\documentclass[a4paper,11pt]{article}

\usepackage{enumitem}
\usepackage{float}
\usepackage{graphicx}
\usepackage{caption}
\usepackage{subcaption}
\usepackage{slashed,nicefrac}

\usepackage{tikz}
\usetikzlibrary{shapes,arrows}
\usetikzlibrary{arrows.meta}
\usetikzlibrary{positioning,fadings}
\usetikzlibrary{decorations.pathmorphing}
\usetikzlibrary{decorations.pathreplacing}
\usetikzlibrary{decorations.markings}
\usetikzlibrary{decorations.pathreplacing,calligraphy}

\usepackage{mathtools}

\usepackage{fullpage}
\usepackage[T1]{fontenc}

\usepackage{amssymb,amsmath,cite}

\usepackage{xcolor,bbold,comment}

\usepackage{amsthm}

\def\be{\begin{equation}\begin{aligned}}
\def\ee{\end{aligned}\end{equation}}

\def\thefootnote{\arabic{footnote}}
\allowdisplaybreaks

\numberwithin{equation}{section}

\begin{document}

{\ }
\vspace{1.6cm}

\begin{center}

\def\thefootnote{\fnsymbol{footnote}}

{\Large \bf 
Goldstino condensation at large N 
}

\vskip 1.5cm

Fotis Farakos \ and \ Matteo Morittu

\vskip 1cm

{\it  Dipartimento di Fisica e Astronomia ``Galileo Galilei''\\ Universit\`a di Padova, Via Marzolo 8, 35131 Padova, Italy}
\vspace{15pt}

\&
\vspace{15pt}

{ \it  INFN, Sezione di Padova \\
Via Marzolo 8, 35131 Padova, Italy}
\vspace{15pt}

\vspace{.5cm}

E-mails: fotios.farakos@pd.infn.it \& matteo.morittu@pd.infn.it 

\vspace{1.3cm}

ABSTRACT 

\end{center}

We analyze the standard fermionic 4D Volkov--Akulov action with N global non-linear supersymmetries. We find that the stationary points of the system are described by an effective potential, written in terms of two composite real scalar fields, which approaches the exact quantum effective potential in the large N limit. We identify the existence of at least two stationary points, one representing the original supersymmetry breaking configuration and the other one corresponding to goldstino condensation, where supersymmetry is restored in the deep IR.

\thispagestyle{empty} 
\setcounter{page}{0}

\baselineskip 6mm

\newpage

\tableofcontents

\section{Introduction}

Fermionic condensation often plays an important role in understanding the vacuum structure of a quantum theory.
One of the basic systems where fermionic condensation has been understood is the Nambu--Jona-Lasinio model \cite{Nambu:1961tp,Nambu:1961fr}, together with its 2D counter-part, the Gross--Neveu model \cite{Gross:1974jv}. (For textbook discussions see e.g. \cite{Coleman:1985rnk,Peskin:1995ev}).
An aspect that makes these models stand out is that, due to a large N number of fermion species, the quantum effective potential can be evaluated with arbitrary precision in a 1/N expansion, and the stationary points can be analyzed with confidence. 

Recently, a new type of fermionic condensation has been brought to the forefront, namely the condensation of the N=1 goldstino \cite{DallAgata:2022abm}.
This effect has a crucial impact on string flux compactifications and signals an intrinsic instability that may be generically present in anti-brane uplifts to de Sitter vacua \cite{Kachru:2003aw,Balasubramanian:2005zx,Conlon:2005ki,Kallosh:2018nrk,Bento:2021nbb,Bena:2022cwb}.
This instability also adds a new obstacle to obtaining 4D long-lived de Sitter critical configurations from supersymmetric string theories \cite{Sethi:2017phn,Gautason:2018gln,Gao:2020xqh,Danielsson:2018ztv,Andriot:2022way} \footnote{Stringy de Sitter vacua that do not make use of non-linear supersymmetry, e.g. \cite{Antoniadis:2018hqy,Antoniadis:2019rkh,Antoniadis:2020stf}, could possibly evade the goldstino condensation instability. However, this is not certain since the effect may still take place within theories with linearly realized supersymmetry. Such a question needs to be addressed separately.}, and may further restrict de Sitter solutions in 4D N=1 supergravity, thus extending some previous results \cite{Cribiori:2020use,Montero:2021otb,Andriot:2021rdy,DallAgata:2021nnr,Emelin:2022wft}.
Furthermore, within N=1 supergravity, the goldstino condensation effect seems to persist \cite{DallAgata:2022abm}, and it should be related to the gravitino condensation, which again shows a tachyonic instability \cite{Jasinschi:1984cx,Ellis:2013zsa,Alexandre:2013iva,Ishikawa:2019pnb,Alexandre:2014lla} \footnote{The condensation of gravitini has also been studied in 4D N=2 supergravity as the sole source for a de Sitter uplift \cite{Kehagias:2009zz}, but with full stability still being an open question.}.

In \cite{DallAgata:2022abm} the existence of the goldstino tachyonic instability was demonstrated within 4D N=1 non-linear supersymmetry by means of the so-called functional renormalization group flow \cite{Polchinski:1983gv}.
The adoption of such a method was necessary because a perturbative loop expansion would not be trustable. 
In general, however, one would like to have at hand a model where the goldstino condensation effect can be understood directly with the use of more conventional methods of quantum field theory. 
In particular, a large N demonstration of the existence of this phenomenon would be welcome.
In the present work we take the first steps towards filling this gap. 
It is of course the 4D N=1 goldstino condensation that has the most phenomenological value, but it is theoretically important to analyze the generic existence of goldstino condensation from various perspectives \footnote{In \cite{Kallosh:2022fsc} a {\it discontinuity} is discussed for the goldstino condensation of \cite{DallAgata:2022abm}. Such discontinuities can indeed appear when condensations take place, but they do not signal an inconsistency \textit{per se}. For example, in the so-called CP$^{N-1}$ model (see e.g. \cite{Coleman:1985rnk}), the original classical critical point is not a critical point of the quantum theory at all, for any finite value of the coupling. Interestingly, the specific bosonic model has a classically spontaneously broken symmetry which is dynamically restored. Then, fields transforming linearly under the restored symmetry are built from the classical Goldstone fields, as it also happens in \cite{DallAgata:2022abm}.}.

We will work with a single and well-established model that includes only fermions: the Volkov--Akulov (VA) model with N non-linearly realized supersymmetries \cite{Volkov:1973ix,Clark:2000rv,Cribiori:2016hdz}.
Since these supersymmetries are non-linearly realized, their number can be arbitrarily large, and the same holds for the number of the accompanying Goldstone fermions, the goldstini.
This means that we are not restricted to the typical N $\leq$ 4 of linearly realized supersymmetry (or N $\leq$ 8 for supergravity).
In addition, our aim here is neither to extract nor to study a phenomenological result, but we would like, instead, to answer one and only one question: can a system with N goldstini have a well-controlled stationary point described by goldstino condensation? Our main result is an affirmative answer to this question, thus giving complementary support to the findings of \cite{DallAgata:2022abm}.
Our results further suggest that such a configuration corresponds to the restoration of supersymmetry, and we also give some arguments in favor of its protection from higher-order corrections.
Finally, we analyze a system that is more relevant for string flux compactifications, where, among N fermions, only one is a 4D N=1 goldstino, with all the others becoming pseudo-goldstini, once they acquire a mass.

\section{N goldstini}

\subsection{The effective action and large N}

We work with a system that has N non-linearly realized supersymmetries and focus explicitly on the goldstino sector, which is described by the Lagrangian
\be
\label{LN1}
{\cal L} = - \frac{{\rm N} f^2}{2} \det\left[A_m{}^a\right] \,, 
\ee
where the goldstino vielbein $A_m{}^a$ is defined as 
\be
\label{A-vielbein}
A_m{}^a = \delta_m^a 
+ \frac{i}{{\rm N} f^2} \sum_{J=1}^{\rm N} \partial_m G_J \sigma^a \overline G^J 
- \frac{i}{{\rm N} f^2} \sum_{I=1}^{\rm N} G_I \sigma^a \partial_m \overline G^I  \,, 
\ee
and $f$ is the supersymmetry breaking order parameter whose mass dimension is $[f]=2$. 
Note that the actual supersymmetry breaking scale is ${\rm N}^{1/4} \sqrt f$. 
The system has a global U(N) R-symmetry under which the spinors $G_I$ and $\overline G^I$ transform in the fundamental and anti-fundamental representation, respectively. The goldstino vielbein $A_m{}^a$ is, instead, a singlet under such U(N).
This theory is defined with a cut-off $\Lambda$ for which we typically assume 
\be
\label{pert}
{\rm N} f^2 > \Lambda^4 \,. 
\ee
For later convenience, and as it is typical in large N models, we have already extracted the N coefficient in front of the starting Lagrangian \eqref{LN1}: the 't Hooft limit \cite{tHooft:1973alw} thus corresponds to ${\rm N} \to \infty$ while keeping $f$ fixed.
Note that, when N is large, both $\sqrt f > \Lambda$ and $\sqrt f < \Lambda$ can satisfy \eqref{pert}.
Let us also observe that the number N of non-linear supersymmetries clearly matches the number of goldstini $I,J = 1, \dots, {\rm N}$, whose transformations are
\begin{eqnarray}
\delta G_{I\,\alpha} = \sqrt{{\rm N}} f \epsilon_{I\,\alpha} + \frac{i}{\sqrt{{\rm N}} f} \sum_{J=1}^{\rm N} 
\left( G_J \sigma^m \, \overline \epsilon^J - \epsilon_J \sigma^m \overline G^J \right) \partial_m G_{I\,\alpha} \,. 
\end{eqnarray}  
Making use of the definition \eqref{A-vielbein}, the leading-order terms of the Lagrangian are
\be
{\cal L} = - \frac{{\rm N} f^2}{2} + i  \sum_{I=1}^{\rm N}  G_I \sigma^m \partial_m \overline G^I + {\cal O}\left(\frac{1}{{\rm N} f^2 }\right) \,.
\ee 
The constant $- {\rm N} f^2 / 2$ can be always removed so that there is no divergent term in the large N limit. 
From now on we will stop inserting the explicit $\sum_{I=1}^{\rm N}$ and summation over the same $I, J$ indices will be implied unless otherwise specified.

For the sake of completeness, note that we are using the conventions of \cite{Peskin:1995ev}: therefore, $\eta_{ab} = {\rm diag}(+1,-1,-1,-1)$, $\sigma^a=(\mathbb{1}_2,\vec{\sigma})$, and, because it will be useful later on, $\gamma^a = \begin{pmatrix} 0 & \sigma^a \\ \overline \sigma^a & 0 \end{pmatrix}$, with $\overline \sigma^a = (\mathbb{1}_2,-\vec{\sigma})$. 

To pursue our aim, namely to get access to stationary points that possibly describe goldstino condensation, we write the theory \eqref{LN1} as 
\be
\label{GaussNG}
{\cal L} = - \frac{{\rm N} f^2}{2} \det[e_m{}^a] 
+ \frac{{\rm N} f^2}{2} C_a{}^m ( e_m{}^a - A_m{}^a ) \,. 
\ee
Once we integrate out $C_a{}^m$, we get 
\be 
\label{CEO}
\frac{\delta {\cal L}}{\delta C_a{}^m} =0  \  \longrightarrow  \   e_m{}^a = A_m{}^a \, 
\ee 
and we recover the model with N goldstini \eqref{LN1}.
Equivalently, the path integration over $C_a{}^m$ yields a delta-function at each point of space-time that enforces the constraint.
The advantage of \eqref{GaussNG} is that the action becomes Gaussian in the fermions and their path integration can be performed, leaving behind only a bosonic theory. In doing so, we will also be able to explicitly demonstrate the large N behavior of the model under consideration.

We start with the path integral 
\be
\label{FPI}
Z = \frac{1}{N_0} \int D[e_m{}^a] D[C_b{}^n] D[G_I] D[\overline G^J] \exp\left[ i \int d^4 x \, {\cal L} \right] \,, 
\ee
where we can split the Lagrangian as ${\cal L} = {\cal L}_B + {\cal L}_F$ with 
\be
\label{MGaussNGB}
{\cal L}_B = - \frac{{\rm N} f^2}{2} \det[e_m{}^a] 
+ \frac{{\rm N} f^2}{2} C_a{}^m ( e_m{}^a - \delta_m{}^a ) \, 
\ee
and 
\be
\label{MGaussNG}
{\cal L}_F = \frac{i}{2} C_a{}^m  
\left( G_I \sigma^a \partial_m \overline G^I - \partial_m G_J \sigma^a \overline G^J \right) \,. 
\ee
The $N_0$ stands for the overall normalization of the path integral. 
To evaluate the fermionic contribution to the path integral we pair the N Weyl goldstini into ${\rm N}/2$ Dirac spinors as follows 
\be
\label{2W1D} 
\Psi^A = \begin{pmatrix} 
G^{(A)}
\\   
\overline G^{(A+{\rm N}/2)} 
\end{pmatrix} \ , \quad A = 1, \dots, {\rm N} / 2 \,.  
\ee
This seems as if we were assuming that N is even. However, we would like to stress that this is just a formality that allows us to easily evaluate the fermion path integral: for odd N the end result would be the same.
Thus, the fermionic contribution to \eqref{FPI} formally reduces to
\be
\label{ZF}
Z_F & = 
\int D[\Psi^A] D[\overline \Psi^A]  \exp \left[ i \int d^4 x \, 
\frac{i}{2} C_a{}^m \sum_{A=1}^{{\rm N}/2}  \left( 
\overline{\Psi}^A \gamma^a \partial_m \Psi^A 
- \partial_m \overline{\Psi}^A \gamma^a \Psi^A 
\right) \right]  
\\[0.5cm] 
& = 
\left( \det[ i C_a{}^m \gamma^a \partial_m ] \right)^{{\rm N}/2} \,, 
\ee
up to the overall $N_0$ factor that we will shortly discuss and specify. 

If we now bring the determinants into the exponential and then into the effective action we get the full bosonic theory 
\be
\label{LN4} 
S_{\rm N} = {\rm N} \times \left\{ - \frac{ f^2}{2} \int d^4 x \left[ \det[e_n{}^b] - C_a{}^m ( e_m{}^a - \delta_m{}^a ) \right]
- \frac{i}{2} \, {\rm tr} \log \left[ i C_a{}^m \gamma^a \partial_m \right] \right\} \,,
\ee
where N crucially appears as a global factor, being at the exponent of the determinant in \eqref{ZF}.
The power of the large N construction is now manifest: by taking N parametrically large we can make the classical effects dominant with arbitrary precision over the quantum effects because higher-loop contributions are always suppressed by N factors compared to the tree-level term. This happens because of the overall N in front of the Lagrangian and means that in the large N limit the quantum effective potential and any stationary points are controlled by the scalar potential of \eqref{LN4}.

Let us notice that the fermion functional determinant contribution is clearly missing a dimensionful normalization inside the logarithm. This is related to the choice of $N_0$ in the path integral normalization, which we should specify. For instance, one can either introduce a scale to match the dimensions or insert the inverse propagator $i \partial \!\!\! /$.
Here we will choose $N_0$ to be 
\be
N_0 = \left( \det\left[ i \partial \!\!\! /\right] \right)^{\frac{{\rm N}}{2}}  \,. 
\ee
Other choices of $N_0$ reflect the different ways that one can use to express the determinant in perturbation theory and should ultimately not matter.

\subsection{The effective potential and stationary points}

Let us now search for stationary points of the bosonic theory \eqref{LN4}. 
From the form of its effective action we directly see that this theory has tensor fields that signal the presence of massive higher-spin excitations \footnote{A preliminary analysis of the spectrum shows that around the original VA point the system has massive excitations, including spin-2 fields, along the lines of \cite{Kraus:2002sa,deRham:2014zqa,Folkerts:2013mra,Ferrara:2018wlb}. Understanding the precise spectrum may be interesting \textit{per se}, but we will not go into the specifics of these excitations here. Let us only note that the interpretation of the excitations should be done with care as pointed out e.g. in \cite{Hawking:2001yt,Ferrara:2018wlb}.}.
However, since we would like to search for translation-invariant and Lorentz-invariant stationary points, we are interested only in the trace parts of these tensors and, consequently, in the resulting scalar potential.
We then describe the VEVs of $C_a{}^m$ and $e_m{}^a$ as follows: 
\be
\label{Cehf} 
C_a{}^m = (1 + h) \, \delta_a{}^m \, , \quad e_m{}^a = (1 + \phi) \, \delta_m{}^a \,. 
\ee
In order to get the full scalar potential we need to reduce the functional determinant that includes $C_a{}^m$ to a convenient form by treating $C_a{}^m$, and so $h$, as a constant {\it background field}.
The reader could be concerned that by ignoring the tensor modes' contribution to the scalar potential we may be missing some non-trivial constraint on the trace parts of the tensors. However, this does not happen for the following reason.
One can think of splitting both $C_a{}^m$ and $e_m{}^a$ into traceful and traceless parts as $C_a{}^m = (1 + h) \, \delta_a{}^m + X_a{}^m$ and $e_m{}^a = (1 + \phi) \, \delta_m{}^a + Y_m{}^a$, where $X_a{}^a\equiv0$ and $Y_a{}^a\equiv0$. Then, it is easy to see that there can never exist a linear term containing either $X_a{}^m$ or $Y_m{}^a$ in the scalar potential simply because there is nothing to contract them with: all terms with $X_a{}^m$ and $Y_m{}^a$ in the potential are directly quadratic in these fields and, therefore, they can be set to vanish consistently when we are searching for a background solution. There can be kinetic mixing of all sorts, of course, but here we are discussing neither the dynamics of the system nor its spectrum.

To proceed, we normalize the fermionic determinant in \eqref{ZF} with $i \partial \!\!\! /$, and treat the bosons as a background. Focusing directly on the relevant contributions from \eqref{Cehf} we can see that \eqref{ZF} becomes
\be
Z_F \ & =  \left(\det\left[ \frac{i C_a{}^m \gamma^a \partial_m }{i \partial \!\!\! /} \right] \right)^{\frac{{\rm N}}{2}} 
 =\left(\det\left[ \frac{i (1+h) \gamma^m \partial_m }{i \gamma^n \partial_n } \right] \right)^{\frac{{\rm N}}{2}} =
 \\ 
& 
 = \left(\det\left[ \frac{- (1+h)^2 \partial^2 \mathbb{1}_4}{- \partial^2 \mathbb{1}_4} \right] \right)^{\frac{{\rm N}}{4}} 
 = \left[ \left(\det\left[ \frac{- (1+h)^2 \partial^2 }{- \partial^2 }\right] \right)^4 
 \times \det[\mathbb{1}_4] \right]^{\frac{{\rm N}}{4}} =
\\ 
& = \left(\det\left[\frac{- (1+h)^2 \partial^2]}{- \partial^2} \right] \right)^{{\rm N}} 
 = \exp\left[ {\rm N} \, {\rm tr} \, \log\left[ \frac{ - (1+h)^2 \partial^2}{ - \partial^2 } \right] \right] \,.
\ee 
We then evaluate the trace of the operator as the sum of its eigenvalues, that is
\be
{\rm tr} \, \log\left[ \frac{ - (1+h)^2 \partial^2}{ - \partial^2 } \right] 
= \sum_k \log\left[ \frac{ (1+h)^2 k^2}{ k^2 } \right]  
= (VT) \int \frac{d^4 k}{(2 \pi)^4} \log\left[(1+h)^2\right]\,, 
\ee
where $VT$ is the four-dimensional volume, which we write as $\int d^4 x$. The expression for the fermionic path integral can thus be brought to the form
\be
\label{IntforVF}
Z_F = \exp\left[ {\rm N} \int d^4 x \int \frac{d^4 k}{(2 \pi)^4}  \log\left[(1+h)^2\right]
\right] \,, 
\ee 
which, once it is evaluated in Euclidean momentum space with a cut-off, gives
\be
\int d^4 x \, V_F(h,\phi) = i \log Z_F = \int d^4 x \, \left( - \frac{{\rm N} \Lambda^4}{32 \pi^2} \log \left[(1+h)^2\right] \right) \,. 
\ee
$V_F$ constitutes the full correction to the bosonic effective potential from the fermion Gaussian integral.
For completeness, in the Appendix we will also discuss such a calculation from the perspective of dimensional regularization.

We conclude that the stationary points of the theory \eqref{LN1} are described by the critical points of the scalar potential 
\be
\label{FullV} 
V_{{\rm eff}} (h,\phi) = {\rm N} \times \left\{ \frac{ f^2}{2} \left[ (1+\phi)^4 - 4 (1 + h) \phi \right]
- \frac{\Lambda^4}{32 \pi^2} \log \left[  (1+h)^2 \right] \right\} \,. 
\ee 
The scalars $h$ and $\phi$ on which $V_{{\rm eff}}$ depends should {\it not} be directly thought of as standard scalar fields, because they are physically parts of the tensor fields $C_a{}^m$ and $e_m{}^a$: $h$ is related to the trace of $C_a{}^m$ and $\phi$ to the trace of $e_m{}^a$.
In spite of this, the critical values of $h$ and $\phi$ do correspond to stationary points of the system and their potential energies at the stationary points can be legitimately identified as actual energy densities \cite{Coleman:1985rnk,Peskin:1995ev}.
The equations for the critical points are 
\be
\label{EQS}
2 f^2 \phi =  - \frac{\Lambda^4}{16 \pi^2 (1 + h)} \, , \quad (1+h) = (1 + \phi)^3  \,,
\ee
and they can be combined into a single equation for $\phi$, which in turn directly gives the value of $h$.
The resultant equation for $\phi$ is 
\be
\phi (1 + \phi)^3 + \frac{\Lambda^4}{32 \pi^2 f^2} = 0 \,. 
\ee
This equation has at least two real solutions that can be easily found if we solve the system numerically, but can also be determined analytically if we solve the equations perturbatively.

Willing to adopt the latter approach, we assume that
\be 
\sqrt f > \Lambda \,. 
\ee 
Under such requirement, we find
\be
\phi_{\rm VA} \simeq -\frac{\Lambda^4}{32\pi^2f^2} + \mathcal{O}\left[ \left(\frac{\Lambda^4}{32\pi^2f^2} \right)^2 \right] \, 
\ee
and 
\be
\label{GoldCond}
\phi_{\rm GC} \simeq -1 + \left(\frac{\Lambda^4}{32\pi^2f^2} \right)^{\frac13} + \frac13 \left(\frac{\Lambda^4}{32\pi^2f^2} \right)^{\frac23} + \frac13 \left(\frac{\Lambda^4}{32\pi^2f^2} \right) + \mathcal{O}\left[ \left(\frac{\Lambda^4}{32\pi^2f^2} \right)^{\frac43} \right] \,. 
\ee
The first solution corresponds to the original VA point and the second one is related to goldstino condensation (GC).
On these critical configurations $h$ takes the values $h = -1 + (1+\phi)^3 \big{|}_{\phi=\phi_{\rm VA}, \phi_{\rm GC}}$.
The potential energies corresponding to the stationary points of \eqref{FullV} that we have just found can then be extracted: they are
\be
{\cal E}_{\rm VA} = \frac{{\rm N} f^2}{2} \left\{ 1 + \mathcal{O}\left[ \left(\frac{\Lambda^4}{32\pi^2f^2} \right)^2 \right] \right\} 
\ee
and
\be
{\cal E}_{\rm GC} = 2 {\rm N} f^2 \left\{\frac{\Lambda^4}{32\pi^2f^2} - \frac{\Lambda^4}{32\pi^2f^2} \log\left[\frac{\Lambda^4}{32\pi^2 f^2}\right] + \mathcal{O}\left[ \left(\frac{\Lambda^4}{32\pi^2f^2} \right)^{\frac43} \right] \right\} \,, 
\ee
respectively.
We see that the configuration $(\phi_{\rm GC},h_{\rm GC})$, which describes the large condensate, has lower energy than the VA point $(\phi_{\rm VA},h_{\rm VA})$ (which seems to correspond to a small condensate; we will shortly come back to this point).
We would like to highlight that, since N is assumed to be very large, the scalar potential \eqref{FullV} is arbitrarily close to the full {\it quantum effective potential}. Therefore, its stationary points $(\phi_{\rm cl},h_{\rm cl})$ (but only those) correspond to actual quantum states of the theory and the corresponding value of the scalar potential captures the energy density of the state with energy $E = (VT) V_{{\rm eff}}(\phi_{\rm cl},h_{\rm cl})$.
As an aside, let us also note that, since the overall value of a potential can always be shifted by a constant, these energy densities have to be considered as {\it relative} one to the other. We will come to this point in a while. 

It is also worth observing that classically, from \eqref{CEO}, we have 
\be
\label{COND}
\phi  = \frac{i}{4 {\rm N} f^2} \left(    \partial_m G_J \sigma^m \overline G^J 
-  G_I \sigma^m \partial_m \overline G^I  \right)  \,. 
\ee
This justifies the interpretation of a non-trivial $\phi$ background value as a signal of goldstino condensation. 
We also see that, when $\phi$ approaches unit, the VEV of the condensate approaches ${\rm N} f^2$ and possibly jeopardises the control over the possible higher-order terms. Indeed, since we are working with the pure VA term \eqref{LN1}, we are certainly ignoring higher-order goldstino self-interactions. 
We will see shortly that, as long as the Lagrangian is written in terms of $A_m{}^a$ and its derivatives, we maintain full control of the vacua at large N. Therefore, no matter what power of $\phi$ we have in the higher-order terms, the condensate is robust even when $\phi$ goes near unit. 

For completeness, let us mention that, contrary to what we have done so far, due to the large number N we could also work under the assumption that $\sqrt f < \Lambda$. After further assuming $\sqrt f \ll \Lambda$, while still respecting ${\rm N} f^2 > \Lambda^4$, in order to extract an analytic result, we find two almost degenerate solutions: $\phi_{\pm} \simeq \pm \frac{\Lambda}{2^{5/4} \sqrt{\pi f}}$ and  $h_{\pm} \simeq \phi_{\pm}^3$. 
The original VA point is not a part of the stationary configurations any more and only solutions corresponding to large condensation correspond to stationary points.
Since the condensate \eqref{COND} takes parametrically large values in this case, we will not pursue this limit further here.

\subsection{The stationary points in the deep IR}

In this subsection we would like to understand what flowing towards the IR means and how $f$ changes while going to lower energies, 
assuming that we are already in a low energy regime where the VA model is weakly coupled, that is 
\be
\label{sqrtfggL} 
\sqrt f \gg \Lambda \,. 
\ee 
Since the VA model has a single coupling, $f$, it would be {\it enough} to evaluate the flow of any specific term or interaction, 
and the other interactions would change accordingly. 
This in principle would require to have a regularization scheme that respects non-linear supersymmetry 
(as, for example, in \cite{DallAgata:2022abm}). 
Within such a setup one could deduce the flow of $f$ by considering the 4-Fermi derivative interaction. 
However, such a calculation would be quite involved for various reasons. 
For instance, 
it would firstly require the identification of a proper regularization scheme, 
and then the evaluation of both the fermionic wave-function renormalization and the actual running of the 4-Fermi vertex. 
Nevertheless, 
because of \eqref{sqrtfggL}, 
one expects that the loop contributions would be subdominant (i.e. ${\cal O}\left(\frac{\Lambda}{\sqrt f}\right)$) with respect 
to the classical running due to the mass dimensionality of $f$. 
Therefore, 
without going into the loop calculation we can focus on the vacuum energy, $V_{\rm vac} = \frac{{\rm N} f^2}{2}$, 
as a tool to infer the classical flow of $f$ towards the IR. 
Even though this is a crude analysis, 
we will not only be able to capture the dominant running, 
but we will also see that the assumption \eqref{sqrtfggL} is enforced by the flow itself. 

Following \cite{Peskin:1995ev}, let us consider the VA model in the form \eqref{LN1} and the Euclidean path integral 
\be
Z = \frac{1}{N_0} \int [D G D \overline{G}]_{\Lambda} \exp \left\{- \int d^4x \, \left[ V_{\rm vac} + i G_I \sigma^m \partial_m \overline{G}^I + \mathcal{O}\left(\frac{1}{N f^2}\right) \right] \right\} \,,
\ee
where
\be
\ [D G D \overline{G}]_{\Lambda} = \prod_{|k|<\Lambda} d G(k) d \overline{G}(k) \,,
\ee
$\Lambda$ representing a momentum cut-off for the quantum field fluctuations (and without specifying the species index, for simplicity).
We then distinguish the integration variables into two groups: the ``high-momentum'' degrees of freedom $(\hat{G}, \hat{\overline{G}})$ that have $b \Lambda \leq |k| < \Lambda$, and the ``low-momentum'' modes $( \tilde{G}, \tilde{\overline{G}})$ carrying a momentum $|k| < b \Lambda$ (the parameter $b$ being a fraction $b < 1$). 
We thus have
\be 
\label{Vvac}
Z = \frac{1}{N_0} 
\int D \tilde{G} D \tilde{\overline{G}} \, e^{ - \int d^4x \left[ V_{\rm vac} + i \tilde{G}_I \sigma^m \partial_m \tilde{\overline{G}}^I + \dots \right]} \int D \hat{G} D \hat{\overline{G}} \, e^{ - \int d^4x \left[ i \hat{G}_I \sigma^m \partial_m \hat{\overline{G}}^I + \dots \right] } 
\,,
\ee
and we further split the path integral normalization factor $N_0$ into the high-momenta and the low-momenta contributions as 
\be
\label{normIR}
N_0 = N_0 \big{|}_{\tilde G} \times N_0 \big{|}_{\hat G} 
= \left(\det\left[ i \sigma^n \partial_n \right]  \right)^{{\rm N}} \Big{|}_{|k| < b \Lambda} 
\times 
\left(\det\left[ i \sigma^n \partial_n \right]  \right)^{{\rm N}} \Big{|}_{b \Lambda \leq |k| < \Lambda} 
\,. 
\ee
Let us now integrate over the high-momentum modes and focus on the vacuum energy change. 
The leading contribution to it comes from the Gaussian kinetic term of $\hat{G}$ and $\hat{\overline{G}}$.
Once we integrate over such degrees of freedom, we find 
\be
\label{deltaVvac}
\Delta V_{\rm vac} \Big{|}_{\rm Gaussian} = 0 \,. 
\ee
Higher-order contributions coming, for instance, from quartic terms of $\hat{G}$ or $\hat{\overline{G}}$ are further suppressed by powers of $f$.
Then, we rescale distances and momenta according to 
\be
\label{b}
x = \frac{x'}{b} \quad \text{ and } \quad k = b k' \,,
\ee
so that the variable $k'$ is still integrated over the range $|k'| < \Lambda$, and the fermionic path integral has once more $[D \tilde{G} D \tilde{\overline{G}}]_{\Lambda}$ as its measure.
Therefore, we have 
\be
\int [D \tilde{G} D \tilde{\overline G}]_{\Lambda} \, e^{- \int d^4x' b^{-4} \left( V_{\rm vac} + \Delta V_{\rm vac} + \dots \right)} 
= 
\int [D \tilde{G} D \tilde{\overline G}]_{\Lambda} \, e^{- \int d^4x' V'_{\rm vac} + \dots} \,, 
\ee
where, in light of \eqref{deltaVvac},
\be 
\label{newVvac}
V'_{\rm vac} = V_{\rm vac} \times \frac{1}{b^4} + \text{ sub-leading contributions} \,. 
\ee  
Because of the rescaling \eqref{b}, while considering \eqref{newVvac}, a decrease of the parameter $b$ represents how much the system flows towards the IR. 
The expression \eqref{newVvac} clearly shows that the vacuum energy tends to increase as $b$ decreases. 
From $V_{\rm vac} = \frac{{\rm N} f^2}{2}$ we deduce that 
\be
\label{fflow}
f = \frac{f_0}{b^2} + \text{ sub-leading contributions} \,, 
\ee
where $f_0$ is the starting value of $f$, before we integrate out any high-momentum modes (i.e. for $b=1$), 
and the sub-leading contributions are of order $\frac{\Lambda}{\sqrt{f_0}} \ll 1$. 
This in turn implies that the coupling accompanying the higher-order interactions of the VA model becomes more and more irrelevant as one flows to the IR. 

We thus conclude that in the deep IR regime $b\to0$ and gives 
\be
\label{DIR}
\frac{\Lambda}{\sqrt f} = \frac{\Lambda}{\sqrt f_0} \times b \to 0 \,. 
\ee 
In such a limit we can check the relative difference between the energy densities of the two critical points: we obtain 
\be
\label{DIRE}
\frac{{\cal E}_{\rm VA} - {\cal E}_{\rm GC}}{{\cal E}_{\rm VA}} \to 1 \quad \text{and} \quad {\cal E}_{\rm VA} \to \frac{{\rm N} f^2}{2} \,. 
\ee
We see that the goldstino condensation point has parametrically lower energy than the VA configuration.
Moreover, under the limit \eqref{DIR} the VA point recovers its classical energy density and all corrections to it vanish.
We see that \eqref{DIRE} is quite suggestive in favor of interpreting the GC point as a supersymmetry restoring field configuration.
Such interpretation is further corroborated by the properties of the kinetic terms of the fermions.
Indeed, in the deep IR regime \eqref{DIR}, where a scalar VEV is properly defined, we observe that 
\be
\phi_{\rm VA} \to 0 \, , \quad h_{\rm VA} \to 0 \,, 
\ee
which means that one recovers the classical stationary point for the VA model, and
\be
\phi_{\rm GC} \to -1 \, , \quad h_{\rm GC} \to -1 \,. 
\ee
Since we know from \eqref{MGaussNG} that the kinetic terms of the fermions on a background defined by the stationary points are 
\be
{\cal L}_{\rm kin} = \frac{i}{2} (1 + h) \left( G_I \sigma^m \partial_m \overline G^I - \partial_m G_J \sigma^m \overline G^J \right) \,, 
\ee
we conclude that at the VA point the fermions have canonical kinetic terms, whereas the kinetic terms of the fermions vanish on the GC point in the deep IR. This absence of appropriate Goldstone modes when $h \to -1$ is consistent with the restoration of supersymmetry \footnote{A similar effect takes place in \cite{Farakos:2020wfc}, where the goldstino stops propagating on the supersymmetric background. One could expect that on such a background the massive spin-2 excitations organize themselves in a supersymmetric way, e.g. along the lines of \cite{Engelbrecht:2022aao}.}. 

As we mentioned earlier, in the Appendix we are going to present the same analysis by using dimensional regularization. 
We will see that in the deep IR regime that we just studied, 
defined by the limit $b\to0$, 
the results from the two different regularization methods nicely match. 

Our findings also connect with the ERG analysis for the 4D N=1 system that used superfields \cite{DallAgata:2022abm}.
There, the system is driven to an asymptotic supersymmetric point where the derivatives of the superpotential vanish, and so does the vacuum energy.
In particular, the asymptotic supersymmetric point of the 4D N=1 system satisfies $G^2 \partial^2 \overline G^2 \sim f^4$, where $G$ is the N=1 goldstino and $\sqrt f$ the N=1 supersymmetry breaking scale. 
We can interestingly observe that \eqref{COND} for $\phi \sim -1$ corresponds to a similar limit. 
For completeness, let us also note that the growth of $f$ in \cite{DallAgata:2022abm} is controlled in the IR by its mass dimension.

\section{Robustness against higher-order terms}

In this section we would like to understand how much the stationary points ($\phi_{\rm VA}, h_{\rm VA}$) and ($\phi_{\rm GC}, h_{\rm GC}$) are influenced by the higher-order terms that our starting model is ignoring.
Even though there are terms that we cannot account for and may change the solutions, especially if they describe R-symmetry breaking, we will provide a simple rule of thumb for the circumstances when higher-order interactions could be dangerous. More precisely: 
\begin{enumerate}[label=(\Alph*)]
\item When higher-order terms appear only through the goldstino vielbein $A_m{}^a$ and its derivatives, the goldstino condensation is {\it always} robust for large N. 
\item When higher-order terms also include explicit $(A^{-1})_a{}^m \partial_m G^I$ terms, the goldstino condensation {\it may} be jeopardized.  
\end{enumerate} 
We will prove (A) and we will give two different examples of (B). 

The reader should keep in mind that, if the goldstino condensation does restore supersymmetry, then higher-order corrections to the treatment above do not threat its existence; only non-perturbative corrections could do that. This is possibly the reason why it is easy to readily control a large class of higher-order corrections of the form (A).
Despite of the lack of a proof, we will also see that the corrections of the form described in (B) seem to remain innocuous most of the time.

\subsection{Corrections from goldstino vielbeins and matter}

Let us start by considering the case in which the higher-order terms are expressed only by the goldstino vielbein $A_m{}^a$ and its derivatives.
Schematically, they have the form 
\be
\label{HOD}
\frac{1}{M^{R-4}} (\partial_n)^R (A_m{}^a)^T 
 \ \longrightarrow  \
\frac{1}{M^{R-4}} (\partial_n)^R (e_m{}^a)^T 
 \ \longrightarrow  \ 
\frac{1}{M^{R-4}} (\partial_n)^R (1 + \phi)^T \,, 
\ee
for some scale $M$ and some powers $R$ and $T$.
Since we are focused on the stationary points of a scalar potential, such derivative terms do not change the outcome.
More importantly, when the higher-order terms take the form \eqref{HOD}, even if they are not only derivative interactions, they are always parametrically sub-leading in the large N limit simply because they have no N factor in front of them.
We can conclude that for large N the goldstino condensation is not spoilt by such higher-order corrections. 

Let us now make the discussion a bit more precise by assuming that we have some massive scalars coupled to the system in a way that preserves the existing non-linear supersymmetry.
We assume that such scalars are in their VEVs so that we can restrict ourselves to consider the Gaussian piece of their action.
These scalar fields could represent some degrees of freedom that have been removed from the spectrum to deduce the low energy goldstino theory. Their impact can serve as a proxy for the higher-order corrections.
We consider $n$ real scalars $b_i$ with 
\be
\Delta {\cal L} = \frac12 \det[e_l{}^c] \eta^{ab} E_a{}^m (\partial_m b_i) E_b{}^n (\partial_n  b_i) 
- \frac12 M^2 \det[e_m{}^a] b_i^2 \ , \quad \text{for } i=1, \dots, n  \,. 
\ee 
We are interested in evaluating the contribution of the functional determinant of the scalars $b_i$ to the effective potential.
To this end, we expand $e_m{}^a$ once more as $e_m{}^a = (1 + \phi) \delta_m{}^a$, treating $\phi$ as a {\it background field}. We get
\be
\int d^4 x \Delta V_{\rm eff} (\phi) & = - \frac{i \, n}{2} \log \det 
\left[ \det[e_l{}^c] \eta^{ab} E_a{}^m \partial_m E_b{}^n \partial_n + M^2 \det[e_m{}^a] \right] = 
\\
\ & = - \frac{i \, n}{2} (VT) \int \frac{d^4 k}{(2 \pi)^4} 
\log \left[- (1+\phi)^2 k^2 + M^2 (1+\phi)^4 \right] 
\,. 
\ee
If we now assume that the $N_0$ for each of the scalar fields $b_i$ corresponds to a free massive scalar, after a Wick rotation, we obtain
\be
\Delta V_{\rm eff}(\phi)= \frac{n}{2} \int \frac{d^4 k_E}{(2 \pi)^4} 
\log \left[\frac{(1+\phi)^2 k_E^2 + M^2 (1+\phi)^4}{k_E^2 + M^2} \right] \,, 
\ee
and, explicitly,
\be
\label{DVEFF}
\Delta V_{\rm eff}(\phi) =  \frac{n}{64 \pi^2} \Big{\{}  & 
\Lambda^4 \log[(1+\phi)^2] 
+ M^2 \Lambda^2 \left[ (1+\phi)^2 -1 \right] +
\\ 
 & +  M^4 \log\left[ \frac{M^2 + \Lambda^2}{M^2} \right] 
 + \Lambda^4 \log\left[ \frac{M^2 (1+\phi)^2 + \Lambda^2}{M^2+\Lambda^2} \right] + 
 \\ 
 & + M^4 (1+\phi)^4   \log\left[ \frac{M^2(1+\phi)^2}{M^2(1+\phi)^2 + \Lambda^2} \right] 
\Big{\}} \,.  
\ee
We would like to see how this new contribution to the effective scalar potential changes the stationary points.
Even though it is not necessary, we can assume that the scalars $b_i$ are heavy, that is $M^2 > \Lambda^2$.
Without actually performing any further calculation, but simply exploiting the large N limit, we observe that the contribution \eqref{DVEFF} to the total scalar potential is parametrically subdominant with respect to \eqref{FullV} as long as 
\be 
{\rm N} \gg n \, . 
\ee 
Indeed, the analysis here falls under the general arguments for the robustness of the stationary points under higher-order deformations of the form \eqref{HOD}.
In particular, as far as \eqref{EQS} is concerned, the left-hand-side equation does not change, whereas the right-hand-side equation becomes 
\be
\label{DEV}
(1+h) = (1+\phi)^3 + \frac{1}{2 {\rm N} f^2} \frac{\partial (\Delta V_{\rm eff}(\phi))}{\partial \phi} \,. 
\ee
From here it is evident that the deviation of \eqref{DEV} from \eqref{EQS} is arbitrarily small at large N.

Let us notice that we can also extend the above conclusion to a more general matter-coupled VA system.
Consider a Lagrangian that has a matter part (made by scalars, vectors, spinors) of the form \cite{Clark:2000rv,Bandos:2016xyu} 
\be
\label{MATC}
{\cal L}_{\rm matter} (A_m{}^a, b, v_{m}, \chi_\alpha) \,. 
\ee
Ultimately, its induced contribution to the quantum effective potential boils down to some $\Delta V_{\rm eff}(\phi)$ and, as a consequence, the deviation from the original system is controlled by \eqref{DEV}, therefore being arbitrarily small at large N. 

We can conclude that goldstino condensation is quite a robust prediction of the large N non-linear supersymmetric theory, assuming that matter is coupled to the starting VA system via \eqref{MATC} (which inevitably preserves also the R-symmetry).

\subsection{Explicit goldstini under derivatives}

We now discuss terms where the goldstini explicitly appear under derivatives,
\be 
D_a G^I = E_a{}^m \partial_m G^I \,, 
\ee 
thus breaking the assumption \eqref{HOD}.
These terms can possibly jeopardize goldstino condensation even at large N. 
For example, one can consider a term like
\be
\label{RBR}
g \, \det[e_m{}^c] \, D_a G^I \sigma^{ab} D_b G^I + \text{c.c.}  \,,
\ee
for some complex dimensionful coupling $g$.
Such a term potentially has a non-trivial impact: not only it is not of the form \eqref{MATC}, but it also contributes to the large N functional determinant because it contains the N goldstini.
However, since we are interested in scalar backgrounds, \eqref{RBR} takes the form 
\be
g \, (1+\phi)^2 \, \partial_a G^I \sigma^{ab} \partial_b G^I + \text{c.c.} \,, 
\ee
and it is then clear that it will never contribute to the quantum effective potential of $\phi$. One can in fact perform an integration by parts, treating $\phi$ as a constant {\it background field} (because we are interested in the properties of the effective potential), to obtain
\be 
\int d^4 x \, g \, (1+\phi)^2 \, \partial_a G^I \sigma^{ab} \partial_b G^I 
 \ \longrightarrow \  
- \int d^4 x \, g \, (1+\phi)^2 \, G^I \sigma^{ab} \partial_a \partial_b G^I  = 0 \,.
\ee 
Let us notice that the same manipulation can be done by going to momentum space and assigning zero momentum to $\phi$, as the standard procedure to evaluate the contributions to the quantum effective potential requires.
We conclude that terms like \eqref{RBR}, if present, do not jeopardize the new stationary point associated to goldstino condensation.

As a further example, we can consider the term 
\be
\label{RBRG}
g'_{IJ} \, \det[e_m{}^c] \, D_a G^I D^a G^J + \text{c.c.}  \,, 
\ee
for some complex dimensionful couplings $g'_{IJ}$.
This term will contribute to the quantum effective potential with a large N coefficient.
However, it manifestly changes the number of degrees of freedom because it leads to $G^I \partial^2 G^J$ terms, which induce an additional massive fermion in the spectrum for each goldstino.
For a consistent EFT such terms should be, in any case, independently highly suppressed.
Even so, let us analyze the impact of \eqref{RBRG}, assuming that one uses it only as an interaction vertex.
The easiest way to handle such a term is to package the goldstini once again into Dirac spinors $\Psi^A$ of the form \eqref{2W1D}, where $A=1,\dots,{\rm N}/2$.
For our analysis we will also assume that the only non-zero contributions to \eqref{RBRG} come from 
\be
g'_{IJ} = g' \times \delta_{_{A,A+{\rm N}/2}} \,, 
\ee 
with $g' \in \mathbb{R}$ now.
Then, considering only the background $h$ and $\phi$ contributions from \eqref{RBRG}, 
the Gaussian fermionic sector is 
\be 
\label{2-DER}
{\cal L} = i (1+h) \overline \Psi^A \gamma^m \partial_m \Psi^A - g' (1+\phi)^2  \overline \Psi^A (i \gamma^m \partial_m)^2 \Psi^A \,, 
\ee
where the first term originates from \eqref{MGaussNG}.
The functional determinant for a single Dirac spinor of \eqref{2-DER} becomes
\be
\det\left[ i (1+h) \partial \!\!\! / - g' (1+\phi)^2  (i \partial \!\!\! /)^2 \right] 
\equiv 
\det\left[ i (1+h) \partial \!\!\! / \right] \times 
\det\left[\mathbb{1}_4 - g' \frac{(1+\phi)^2}{1+h}  i \partial \!\!\! / \right] \,.
\ee
We readily see why such a deformation changes the degrees of freedom and introduces new massive fermions.
However, as far as our purpose is concerned, we simply need to treat the new contribution to the effective potential from the new massive fermionic functional determinant, having in mind that $\det\left[ i (1+h) \partial \!\!\! / \right]$ is already included in the effective potential and corresponds to the original goldstini.
One way to do this calculation is to recast the overall functional determinant in the form 
\be
\det\left[ i (1+h) \partial \!\!\! / - g' (1+\phi)^2  (i \partial \!\!\! /)^2 \right] 
\equiv 
\det\left[ - g' (1+\phi)^2 i \partial \!\!\! / \right] \times 
\det\left[ i \partial \!\!\! / - \frac{(1+h)}{g' (1+\phi)^2} \mathbb{1}_4 \right] \,.
\ee
The first term is similar to that which we have already calculated, but with $1+h$ replaced by $- g' (1+\phi)^2$. The second factor corresponds, instead, to the contribution of a massive fermion with canonical kinetic term \footnote{To evaluate the contribution of the massive fermions we notice that $\det\left[ i \partial \!\!\! / - m \mathbb{1}_4 \right] 
= \det\left[(i \partial \!\!\! / - m \mathbb{1}_4 ) \gamma_5^2 \right] 
= \det\left[(i \partial \!\!\! / - m \mathbb{1}_4 ) \gamma_5 \right] \det[\gamma_5] 
= \det\left[\gamma_5 (i \partial \!\!\! / - m \mathbb{1}_4 ) \gamma_5 \right] = \det\left[-i \partial \!\!\! / - m \mathbb{1}_4 \right]$. 
This allows us to express the functional determinants of massive Dirac fermions in terms of functional determinants of massive scalars, 
giving the known result: $\det\left[ i \partial \!\!\! / - m \mathbb{1}_4 \right] 
= \left(\det\left[ \partial^2 + m^2\right]\right)^2 $.}. 
We conclude that the potential that we have to extremize is (up to constants) 
\be
\label{FullV-X} 
V_{\rm eff} (h,\phi) = & \frac{{\rm N}  f^2}{2}  \left[ (1+\phi)^4 - 4 (1 + h) \phi \right]
- \frac{{\rm N} \Lambda^4}{32 \pi^2} \log \left[ (1+\phi)^4 \right] +
\\
\ & - \frac{{\rm N} m^4 }{32 \pi^2} 
\left( 
 \frac{\Lambda^2}{m^2} 
+ \log\left[ 
\frac{m^2}{m^2 + \Lambda^2}  
\right]
- \frac{\Lambda^4}{m^4} \log\left[ 
\frac{\Lambda^2}{m^2 + \Lambda^2}  
\right]
\right) 
\,, 
\ee 
where 
\be
m = \frac{(1+h)}{g' (1+\phi)^2} \,. 
\ee
Contrary to the previous case, we cannot use the large N limit any more to eliminate the new terms. However, the theory still has a valid large N limit and the stationary points of \eqref{FullV-X} correspond to stationary points of the full quantum effective potential (in such a limit). 

We would like to investigate the existence of a goldstino condensate for the scalar potential \eqref{FullV-X}. In order to be able to continue analytically we make the assumption that there is a hierarchy between the scales at play, namely 
\be
\label{MggL} 
M \gg \Lambda \, , \quad \text{once } \, M \equiv \frac{1}{g'} \,, 
\ee 
and we furthermore assume that 
\be
\label{mggL}
m = \frac{M (1+h)}{(1+\phi)^2} \gg \Lambda \,. 
\ee
We will check that this condition holds on the solutions. We can already see that it is satisfied for the VA point, if the latter persists.
Under such assumption, and up to constants, the potential that we need to extremize takes exactly the form \eqref{FullV} at leading order in the $\Lambda^2/m^2$-expansion.
In particular, the corrections are all of the form $\Lambda^4 \times (\Lambda^2/m^2 + \Lambda^4/m^4 + \dots)$.
Therefore, both the VA configuration and the GC solution remain intact. 
Finally, one can check that both solutions satisfy \eqref{mggL}, 
as long as 
\be
M^3 \Lambda \gg f^2 \,. 
\ee
This further implies that $M \gg \sqrt f$. 

\begin{table}
\begin{center}
\begin{tabular}{||c c | c c c | c c ||} 
 \hline
 $f$ & $M$ &  $\phi_{\rm GC}$ & $h_{\rm GC}$ & ${\cal E}_{\rm GC}/$N & $\phi_{\rm VA}$ & $h_{\rm VA}$ \\ [0.5ex] 
 \hline\hline
 $591$ & $17.1 \! \times \! 10^{3}$ 
 & $-1+10^{-2}$ & $-1 + 4 \! \times \! 10^{-9}$ & $6 \! \times \! 10^{-2}$ 
 & $-3 \! \times \! 10^{-8}$ & $-6 \! \times \! 10^{-8}$ \\ 
 \hline
 $24$ & $670$ 
 & $-1+5 \! \times \! 10^{-2}$ & $-1 + 3 \! \times \! 10^{-6}$ & $4 \! \times \! 10^{-2}$ 
 & $-6 \! \times \! 10^{-6}$ & $-2 \! \times \! 10^{-5}$ \\
 \hline
 $6$ & $162$ 
 & $-1+10^{-1}$ & $-1 + 4 \! \times \! 10^{-5}$ & $3 \! \times \! 10^{-2}$ 
 & $-9 \! \times \! 10^{-5}$ & $-3 \! \times \! 10^{-4}$ \\
 \hline
\end{tabular} 
\caption{\label{TAB1} \small \it 
Few instances of numerical solutions for stationary points of \eqref{FullV-X} with $\Lambda=1$, but without making any approximation on the effective potential. The numerical solutions approach the analytic ones as we go closer to the parametric limits that allow our approximations.
The vacuum energy at the VA point is always in very good agreement with ${\rm N} f^2/2$ and therefore we do not write it explicitly.
Note that, because we are interested in the orders of magnitude and in the possible existence of a solution, we have rounded-up the presented numerical results.} 
\end{center}
\end{table}

Let us observe that we have worked with a parametric separation between the various scales ($M$, $\sqrt f$ and $\Lambda$) that enter the problem so that we can easily deduce analytic results.
Clearly, the solutions still exist for weaker assumptions but they have to be found numerically: we provide few numerical solutions for more conservative values of the coefficients in Table \ref{TAB1}.
We do not know under which conditions the solutions will seize to exist and if they seize to exist at all.
When $M$ becomes smaller than $\sqrt f$, the extremization problem can not be approached easily by the adoption of analytical methods, and also the numerical analysis seems to require stronger machines or more refined techniques. 

We conclude that higher-order terms with explicit derivatives of the goldstini may seem harmful at first sight, but it is not obvious that they actually have an impact on the system after all.
As the reader has appreciated, we have analyzed few such terms and we have seen that the properties of the stationary points do not considerably change. Nonetheless, we do not have a general argument to state that the higher-order corrections lying under the circumstance (B) cannot threat the goldstino condensate. As an aside final remark, let us note that other higher-order terms of a similar form can in principle be reduced to the Gaussian terms that we have studied by using Lagrange multipliers.

\section{A single goldstino and N-1 pseudo-goldstini}

In this section we wish to take advantage of the large number of fermions in the system to deduce a result for a model that has only N=1 non-linear supersymmetry.
To do this, we make all the fermions massive but one, which corresponds to the single goldstino that the theory has in the low energy regime. 

We split the goldstini as 
\be
G^I = ( G^0 , G^i ) \,, 
\ee
where $G^0$, which we will denote as $G$ from now on, represents the goldstino for the N=1 non-linear supersymmetry, and the $G^i$s are 2n pseudo-goldstini for a reason that will be clarified in a while.
For convenience we pair the 2n pseudo-goldstini into n Dirac spinors $\Psi^A$ following \eqref{2W1D}. We then explicitly break the extended non-linear supersymmetries down to 1 by introducing a Dirac mass term 
\be
M \det[A_m{}^a] \overline{\Psi}^A \Psi^A \,. 
\ee 
Note that, if each Dirac spinor is split into two Majorana fermions, then one gets a Majorana mass term for each one of them.
Since we get back the full N goldstini system when these masses vanish, we call the massive fermions $\Psi^A$ pseudo-goldstini.
We have
\be
{\cal L} = - \frac{{\rm N} f^2}{2} \det[e_m{}^a] 
+ \frac{{\rm N} f^2}{2} C_a{}^m ( e_m{}^a - A_m{}^a) 
+  M \det[e_m{}^a] \overline{\Psi}^A \Psi^A \,, 
\ee
with 
\be
A_m{}^a = \delta_m{}^a 
+ \frac{i}{{\rm N} f^2}  \partial_m G \sigma^a \overline G  
- \frac{i}{{\rm N} f^2} G \sigma^a \partial_m \overline G 
- \frac{i}{{\rm N} f^2}  \overline{\Psi}^A \gamma^a \partial_m \Psi^A 
+  \frac{i}{{\rm N} f^2}  \partial_m \overline{\Psi}^A \gamma^a \Psi^A \,. 
\ee 
As elsewhere throughout this paper, we are interested in stationary points that are translation-invariant and Lorentz-invariant: we will consider directly the trace parts of $C_a{}^m$ and $e_m{}^a$ as in \eqref{Cehf}. 

We want once more to perform the Gaussian integral over the fermions and derive the contribution to the effective potential for $h$ and $\phi$.
To deduce the relevant modifications to it we perform two formal steps that allow us to get the result directly from the formulas that we already have at our disposal.
First, we redefine all the fermions as follows 
\be
\label{Gredef}
G^I \ \longrightarrow \ \frac{1}{\sqrt{1+h}} G^I \,, 
\ee
treating $h$, as always, as a constant (because we are only interested in the effective potential critical points).
As a consequence, the Lagrangian takes the form 
\be
{\cal L} = & - \frac{{\rm N} f^2}{2} (1+\phi)^4 
+ \frac{{\rm N} f^2}{2} 4 \phi  (1+h) 
+  \frac{M (1+\phi)^4}{1+h}  \overline{\Psi}^A \Psi^A  +
\\ 
\ & 
- \frac{i}{2} ( \partial_m G \sigma^m \overline G - G \sigma^m \partial_m \overline G 
- \overline{\Psi}^A \gamma^m \partial_m \Psi^A + \partial_m \overline{\Psi}^A \gamma^m \Psi^A) +
\\ 
\ & 
+ \frac{{\rm N} \Lambda^4}{32 \pi^2} 
\log \left[ 
 (1+h)^2 
\right] 
\,. 
\ee
The last term appears from the fermionic measure in the path integral because of the redefinition of the fermions. It has to be so in light of the fact that, if no field redefinition is performed, such a contribution appears from the Gaussian integral over the massless fermions. After \eqref{Gredef} the massless fermion $G$ decouples and it can be eliminated without any effect. This implies that the fermion redefinition has to contribute to the Lagrangian through the path integral measure.
Before proceeding, as a reminder for the reader, let us note that ${\rm N} = 2 {\rm n} + 1$. 

As we were mentioning just above, the goldstino $G$ can now be integrated over without any effect, except of an overall shift in the vacuum energy, which we ignore. After eliminating $G$ right away, the remaining 2n fermions are combined into n massive Dirac spinors with canonical kinetic terms and mass $M (1+\phi)^4 / (1+h)$.
Knowing from the previous section how to evaluate the functional integral of these massive fermions, once we integrate out all the fermions by performing the corresponding Gaussian integral, we explicitly find 
\be
\label{FullV-PG} 
V_{\rm eff} (h,\phi) = & \frac{{\rm N}  f^2}{2}  \left[ (1+\phi)^4 - 4 (1 + h) \phi \right]
- \frac{{\rm N} \Lambda^4}{32 \pi^2} \log \left[ (1+h)^2 \right] +
\\
\ & - \frac{({\rm N}-1) m^4 }{32 \pi^2} 
\left( 
 \frac{\Lambda^2}{m^2} 
+ \log\left[ 
\frac{m^2}{m^2 + \Lambda^2}  
\right]
- \frac{\Lambda^4}{m^4} \log\left[ 
\frac{\Lambda^2}{m^2 + \Lambda^2}  
\right]
\right) 
\,, 
\ee 
where 
\be
m = \frac{M (1+\phi)^4}{1+h}  \,. 
\ee
We see that the large N limit still gives us a reliable approximation to the full quantum effective potential with arbitrary precision. 
To derive the stationary points we extremize \eqref{FullV-PG} with respect to $h$ and $\phi$. 
To avoid clutter we also assume that in the large N limit we can have ${\rm N} \simeq {\rm N} -1$. 
Then, combining the equations for $h$ and $\phi$ we see that we can readily eliminate $h$ because it is bound to satisfy 
\be
\label{1+h-PG}
1 + h = \frac{8 \pi^2 f^2 (1+\phi)^4 - \Lambda^4}{8 \pi^2 f^2 (1 + 5 \phi)} \,. 
\ee
The system of the equations of extremization of \eqref{FullV-PG} therefore reduces to a single equation for $\phi$. 
It is possible to search for stationary point solutions without making any assumption on a hierarchy among the various scales at work, but, in this case, one has to proceed numerically. For completeness, we give some numerical results in Table \ref{TAB2}. 

However, we can easily proceed analytically by first invoking the typical hierarchy $\Lambda \ll \sqrt{f}$ and consequently observing that, under such requirement, \eqref{1+h-PG}, for the goldstino condensation point, gives 
\be
1 + h_{\rm GC} \simeq  \frac{\Lambda^4}{32\pi^2f^2} \,, 
\ee
assuming that
\be
\phi_{\rm GC} \simeq - 1  \, , \quad f^2 (1+\phi_{\rm GC})^4 \ll \Lambda^4\,. 
\ee
These equations are in complete agreement with \eqref{GoldCond} and we will also verify them on the solution. 
Always within such limits the equation for $\phi_{\rm GC}$ is
\be
2 & f^2 (1 + \phi)^3 - \frac{\Lambda^4}{16 \pi^2} 
- \Lambda^4 \frac{512 f^4 M^2 \pi^2 (1 + \phi)^7}{\Lambda^{10}} =
\\
= & \, - 2 (1 + \phi) \pi^2 \Lambda^{4}  \left(\frac{512 f^4 M^2 \pi^2 (1 + \phi)^7}{\Lambda^{10}} \right)^2
\log\left[ 1 + \frac{1}{2 (1 + \phi) \pi^2} \frac{\Lambda^{10}}{512 f^4 M^2 \pi^2 (1 + \phi)^7} \right] \,. 
\ee
From this relation we can recover the goldstino condensation solution \eqref{GoldCond}, if we assume that 
\be
\label{fMllL}
\frac{512 f^4 M^2 \pi^2 (1 + \phi)^7}{\Lambda^{10}} \ll 1 \,, 
\ee
which (for \eqref{GoldCond}) reduces to $M^3 \ll \Lambda f$ and therefore also to $M^2 \ll f$. 
Consistently with these bounds we can still have $M>\Lambda$ or $M<\Lambda$: the original goldstino condensation solution is intact for arbitrarily light or for quite heavy pseudo-goldstini. 
The VA point also remains intact when we have $\sqrt f \gg \Lambda \gg M$. 
It is of course good that the limit $M \to 0$ can be taken smoothly: in this limit the N-1 pseudo-goldstini 
become goldstini and we recover the results for the original N goldstini model of Section 2.

\begin{table}
\begin{center}
\begin{tabular}{||c c | c c c | c c ||} 
 \hline
$f$ & $M$ &  $\phi_{\rm GC}$ & $h_{\rm GC}$ & ${\cal E}_{\rm GC}/$N & $\phi_{\rm VA}$ & $h_{\rm VA}$ \\ [0.5ex] 
 \hline\hline 
 $8.8 \! \times \! 10^6$ & $0.16$ 
 & $-1 + 10^{-4}$ & $-1 + 2 \! \times \! 10^{-17}$ & $0.25$  
 & $-4 \! \times \! 10^{-17}$ & $- 10^{-16}$ \\ 
 \hline
 $4595$ & $4$ 
 & $-1 + 10^{-3}$ & $-1 + 2 \! \times \! 10^{-10}$ & $0.15$  
 & $-6 \! \times \! 10^{-12}$ & $-6 \! \times \! 10^{-10}$ \\
 \hline
 $1122$ & $0.017$ 
 & $-1 + 10^{-2}$ & $-1 + 2 \! \times \! 10^{-11}$ & $0.15$  
 & $-3 \! \times \! 10^{-9}$ & $-8 \! \times \! 10^{-9}$ \\ 
 \hline
\end{tabular} 
\caption{\label{TAB2} \small \it 
Few instances of numerical solutions for stationary points of \eqref{FullV-PG} with $\Lambda=1$, but without making any approximation at the level of the effective potential. 
The numerical solutions approach the analytic solutions as we go closer to the parametric limits that allow our approximations. 
Since the vacuum energy at the VA point is always in very good agreement with ${\rm N} f^2/2$, we do not write it explicitly. 
Here again, because we are interested only in the orders of magnitude and in the existence of a solution, we have rounded-up the presented numerical results.} 
\end{center}
\end{table}

This result is relevant for string flux compactifications that include anti-D3/O3 systems as, for example, KKLT does (where supersymmetry is non-linear \cite{Lindstrom:1979kq,Farakos:2013ih,Kapustnikov:1981de,Bergshoeff:2015tra,Dudas:2015eha,Bergshoeff:2015jxa,Kallosh:2014wsa,Dasgupta:2016prs,Vercnocke:2016fbt,Kallosh:2016aep,Cribiori:2019hod,DallAgata:2016syy}). 
For the sake of the discussion, let us extrapolate our large N results to the case where ${\rm N}=4$.
In \cite{Bergshoeff:2015jxa} the masses of the extra fermions living on the anti-brane world-volume are discussed, in particular for the three massive fermions belonging to the anti-brane, which are pseudo-goldstini. The mass of these fermions is determined by the $n^{(2,1)}$ ISD flux.
As we have seen here a small $M$ or, equivalently, a small $n^{(2,1)}$ ISD flux may lead to goldstino condensation and further support the existence of such effects on an anti-brane. Conversely, if $n^{(2,1)}$ is large, then $M$ becomes large as well, and the system has further issues due to large tadpoles \cite{Bena:2020xrh,Plauschinn:2021hkp}.

\section{Discussion}

In this work we have investigated the existence of new stationary points in the standard 4D Volkov--Akulov fermionic system in the presence of N non-linear supersymmetries.
An intuitive way to think of such an investigation and of our findings is the following.
From the standard Lagrangian $\det [A_m{}^a]$ describing the VA system (see \eqref{LN1}), one can derive the classical equations of motion 
\be
\det [A_m^a] (A^{-1})_b{}^n \sigma^b \partial_n G = 0  \,, 
\ee
and suspect that these equations have two types of vacuum solutions:
\be
\langle G_\alpha \rangle = 0 \quad \text{or} \quad \langle \det [A_m{}^a] \rangle = 0 \,. 
\ee
Clearly, the vacuum solution where the goldstino vanishes corresponds to the original VA point that describes supersymmetry breaking: there, in fact, $\langle \det [A_m{}^a] \rangle = 1$.
The solution where the goldstino vielbein determinant vanishes, instead, corresponds to a condensation of the goldstini (see \eqref{A-vielbein} for the form of $A_m{}^a$), and implies that supersymmetry is restored, because the vacuum energy is now vanishing.
The actual computation is more involved than simply solving $\langle \det [A_m{}^a] \rangle = 0$ for the goldstini and proceeds with path integral methods \cite{Coleman:1985rnk,Peskin:1995ev}, which allow to properly treat fermionic condensates.
However, the naive intuitive expectation turns out to be correct and a solution of the form $\langle \det [A_m{}^a] \rangle = 0$ does actually exist, as the path integral method that we followed here verifies. 

As we already mentioned in the Introduction, even though in the present paper we are working directly in the component form and we are exploiting large N methods, our results lend further support to the goldstino condensation analysis of \cite{DallAgata:2022abm} that was performed with the ERG technique for superfields.
These two approaches can be considered complementary and it is gratifying to see that they agree.
It is also important to bring to the reader's attention the fact that bosonic systems with bosonic Goldstone modes can have a similar behavior where the classically broken symmetry is restored by quantum effects (\cite{Schnitzer:1974ji,Coleman:1974jh,Kobayashi:1975ev,Abbott:1975bn}; see \cite{Coleman:1985rnk,Peskin:1995ev} for textbook analysis).
The fact that something similar happens for fermionic systems should not come as a big surprise, then.
Moreover, this does not mean that supersymmetry cannot be broken, but it signals that the breaking of supersymmetry is more intricate than what one naively expects and has to be studied with care.

It is also worth noting that our results give further evidence that the anti-D3-brane/O3-plane system is inherently unstable on a flux-less Minkowski background: such a system corresponds, in fact, to N=4 \cite{Kallosh:2014wsa}. 
It is true that N=4 is not large, but the large N results may still persist. In this respect, evaluating the leading 1/N corrections to the potential will be illuminating. 

We have also discussed a system where all but one goldstini get masses. 
Such a setup corresponds to placing the anti-D3-brane/O3-plane system on a flux background \cite{Bergshoeff:2015jxa}.
In this setup we have seen that the goldstino condensation persists. 
Such a model can also be studied with the use of constrained superfields satisfying $X^2=0=XY^i$ \cite{Vercnocke:2016fbt}, and exploiting the ERG technique to analyze the existence of condensates.
The resulting (supersymmetric) backgrounds corresponding to the condensation may be ultimately related to some kind of brane-flux annihilation \cite{Kachru:2002gs,Gautason:2015tla}, but we cannot know if this is indeed the case yet. This is one of the important questions that we leave for future studies.
In addition, the impact of including gravitation in our analysis is not necessarily trivial. 

Another path that deserves to be investigated is how the goldstino condensation behaves in different dimensions.
It is worth performing a similar analysis for example in 2D or in 3D, especially taking into account that spin-2 fields and gauge fields behave differently compared to the 4D case.
An analysis of the condensate directly in 10D would also be illuminating, and especially interesting for the BSB models \cite{Mourad:2017rrl}. However, such a study seems more challenging compared to that of the lower dimensional systems.

\section*{Acknowledgements} 

We thank Maxim Emelin, Cristiano Germani, Alfredo Walter Guerrera, Gabriele Levati, Luca Martucci, Stefano Di Noi and Rikard von Unge for helpful comments and discussions. The work of F. F. is supported by the MIUR-PRIN contract 2017CC72MK003.

\appendix

\section{Dimensional regularization and stationary points}

The reader may ask what happens if we utilized dimensional regularization when evaluating, for instance, the momentum integral of \eqref{IntforVF} in Section 2.
Here we are then going to work out the pure large N Volkov--Akulov model and the corresponding integrals by means of dimensional regularization (instead of using cut-off regularization as we did in the bulk of the paper).
Once done, we will compare the results.

We directly consider the calculation of the relevant integral for \eqref{IntforVF}, that is 
\be
\label{dimregint}
i  {\rm N} \log [(1+h)^2 ]  \int \frac{d^4 k}{(2 \pi)^4} \,. 
\ee
The cut-off prescription gives $\int \frac{d^4 k}{(2 \pi)^4} = \frac{i\Lambda^4}{32 \pi^2}$, whereas the integral vanishes within dimensional regularization.
Let us then evaluate the integral  
\be
\int \frac{d^4 k}{(2 \pi)^4} \frac{(k^2)^2}{(k^2 - M^2)^2} \,, 
\ee
using dimensional regularization and taking, only in the very end, the limit $M \to 0$ to make contact to \eqref{dimregint}.
We find 
\be
\mu^{2\epsilon} \int \frac{d^d k}{(2 \pi)^d} \frac{(k^2)^2}{(k^2 - M^2)^2} = \frac{3 i M^4}{8 \pi^2} \Gamma[-2 + \epsilon] \left( \frac{4 \pi \mu^2}{M^2} \right)^{\epsilon} \,, 
\ee
where $d = 4 - 2 \epsilon$.
When sending $M \to 0$, we recover the known result that the integral \eqref{dimregint} vanishes within dimensional regularization.

If we went through all the analysis that we did in the bulk of the paper, we would find that 
\be
\phi_{\rm VA} = - \frac{3 M^4 \Gamma[-2+\epsilon]}{8 \pi^2 f^2} \left( \frac{4\pi\mu^2}{M^2} \right)^{\epsilon} + \dots \, , \quad 
\phi_{\rm GC} = -1 + \left[ \frac{3 M^4 \Gamma[-2+\epsilon]}{8 \pi^2 f^2} \left( \frac{4\pi\mu^2}{M^2} \right)^{\epsilon} \right]^{\frac13} + \dots \,,
\ee
the first solution corresponding to the original VA point and the second one to the goldstino condensation configuration.
As far the energy densities of the stationary points are concerned, we have
\be
{\cal E}_{\rm VA} = \frac{{\rm N} f^2}{2} \left\{ 1 + 6 \left[ \frac{3 M^4 \Gamma[-2+\epsilon]}{8 \pi^2 f^2} \left( \frac{4\pi\mu^2}{M^2} \right)^{\epsilon} \right]^2  + \dots \right\}\,
\ee
and 
\be 
{\cal E}_{\rm GC} = \frac{3 {\rm N} M^4 \Gamma[-2+\epsilon]}{4 \pi^2} \left( \frac{4\pi\mu^2}{M^2} \right)^{\epsilon} \left\{1 - \log\left[ \frac{3 M^4 \Gamma[-2+\epsilon]}{8 \pi^2 f^2} \left( \frac{4\pi\mu^2}{M^2} \right)^{\epsilon} \right] + \dots \right\}  \,. 
\ee
Let us now send $M \to 0$. We obtain that
\be
\text{Original VA point:} \quad \phi_{\rm VA} \to 0 \, , \quad {\cal E}_{\rm VA} \to \frac{{\rm N} f^2}{2} \,, 
\ee
and 
\be
\text{Goldstino condensation configuration:} \quad \phi_{\rm GC} \to -1 \, , \quad  {\cal E}_{\rm GC} \to 0 \,. 
\ee
A few comments are in order.
First of all, any dependence on the regularization scheme has dropped out. 
In addition, the VA stationary point has the original vacuum energy value and there is no condensate appearing at that point. In the new stationary point the condensate reaches its maximum value which is independent of the regularization, namely $\phi_{\rm GC} \to -1$, while its energy density vanishes.
Due to the exact vanishing of the vacuum energy we can deduce, giving further support to what we state in the bulk, that supersymmetry has to be restored at that point.
As we already mentioned, let us notice also that the goldstini stop propagating in such a limit at the goldstino condensation point.

Even though, because of the freedom to shift energies, the true value of the energy density in a QFT is a relative matter, we see that the original VA stationary point reaches its original energy density in the limit $M \to 0$.
We can therefore define the energy density of the supersymmetry breaking point with respect to that limit as
\be
\rho_{\rm VA} = \langle {\rm VA} |  |Q|^2 | {\rm VA} \rangle = \frac{{\rm N} f^2}{2} \, ; 
\ee
and in the same limit we also find
\be
\rho_{\rm GC} = \langle {\rm GC} |  |Q|^2 | {\rm GC} \rangle  = 0  \ \longrightarrow \ Q | {\rm GC} \rangle  = 0 \,,
\ee
thus interpreting the GC point as a supersymmetry restoration point.
We see that this analysis agrees exactly with the analysis that we did in Section 2 for the deep IR limit.
It is gratifying to see that the results do not change depending on the regularization scheme. For this reason we work only with the cut-off regularization prescription in the bulk of the article.


\begin{thebibliography}{99}


\bibitem{Nambu:1961tp}
Y.~Nambu and G.~Jona-Lasinio,
``Dynamical Model of Elementary Particles Based on an Analogy with Superconductivity. 1.'',
Phys. Rev. \textbf{122} (1961), 345-358


\bibitem{Nambu:1961fr}
Y.~Nambu and G.~Jona-Lasinio,
``DYNAMICAL MODEL OF ELEMENTARY PARTICLES BASED ON AN ANALOGY WITH SUPERCONDUCTIVITY. II'',
Phys. Rev. \textbf{124} (1961), 246-254


\bibitem{Gross:1974jv}
D.~J.~Gross and A.~Neveu,
``Dynamical Symmetry Breaking in Asymptotically Free Field Theories'',
Phys. Rev. D \textbf{10} (1974), 3235


\bibitem{Coleman:1985rnk}
S.~Coleman,
``Aspects of Symmetry: Selected Erice Lectures'',
Cambridge University Press, 1985. 


\bibitem{Peskin:1995ev}
M.~E.~Peskin and D.~V.~Schroeder,
``An Introduction to quantum field theory'', 
Addison-Wesley (1995).  


\bibitem{DallAgata:2022abm}
G.~Dall'Agata, M.~Emelin, F.~Farakos and M.~Morittu,
``Anti-brane uplift instability from goldstino condensation'',
JHEP \textbf{08} (2022) 005
[arXiv:2203.12636 [hep-th]].


\bibitem{Kachru:2003aw}
S.~Kachru, R.~Kallosh, A.~D.~Linde and S.~P.~Trivedi,
``De Sitter vacua in string theory'',
Phys. Rev. D \textbf{68} (2003), 046005
[arXiv:hep-th/0301240 [hep-th]].


\bibitem{Balasubramanian:2005zx}
V.~Balasubramanian, P.~Berglund, J.~P.~Conlon and F.~Quevedo,
``Systematics of moduli stabilisation in Calabi-Yau flux compactifications'',
JHEP \textbf{03} (2005), 007
[arXiv:hep-th/0502058 [hep-th]].


\bibitem{Conlon:2005ki}
J.~P.~Conlon, F.~Quevedo and K.~Suruliz,
``Large-volume flux compactifications: Moduli spectrum and D3/D7 soft supersymmetry breaking'',
JHEP \textbf{08} (2005), 007
[arXiv:hep-th/0505076 [hep-th]].


\bibitem{Kallosh:2018nrk}
R.~Kallosh and T.~Wrase,
``dS Supergravity from 10d'',
Fortsch. Phys. \textbf{67} (2019) no.1-2, 1800071
[arXiv:1808.09427 [hep-th]].


\bibitem{Bento:2021nbb}
B.~V.~Bento, D.~Chakraborty, S.~L.~Parameswaran and I.~Zavala,
``A new de Sitter solution with a weakly warped deformed conifold'',
JHEP \textbf{12} (2021), 124
[arXiv:2105.03370 [hep-th]].


\bibitem{Bena:2022cwb}
I.~Bena, E.~Dudas, M.~Gra\~na, G.~L.~Monaco and D.~Toulikas,
``Bare-Bones de Sitter'',
[arXiv:2202.02327 [hep-th]].


\bibitem{Sethi:2017phn}
S.~Sethi,
``Supersymmetry Breaking by Fluxes'',
JHEP \textbf{10} (2018), 022
[arXiv:1709.03554 [hep-th]].


\bibitem{Danielsson:2018ztv}
U.~H.~Danielsson and T.~Van Riet,
``What if string theory has no de Sitter vacua?'',
Int. J. Mod. Phys. D \textbf{27} (2018) no.12, 1830007
[arXiv:1804.01120 [hep-th]].


\bibitem{Gautason:2018gln}
F.~F.~Gautason, V.~Van Hemelryck and T.~Van Riet,
``The Tension between 10D Supergravity and dS Uplifts'',
Fortsch. Phys. \textbf{67} (2019) no.1-2, 1800091
[arXiv:1810.08518 [hep-th]].


\bibitem{Gao:2020xqh}
X.~Gao, A.~Hebecker and D.~Junghans,
``Control issues of KKLT'',
Fortsch. Phys. \textbf{68} (2020), 2000089
[arXiv:2009.03914 [hep-th]].


\bibitem{Andriot:2022way}
D.~Andriot, L.~Horer and P.~Marconnet,
``Charting the landscape of (anti-) de Sitter and Minkowski solutions of 10d supergravities'',
JHEP \textbf{06} (2022), 131
[arXiv:2201.04152 [hep-th]].


\bibitem{Antoniadis:2018hqy}
I.~Antoniadis, Y.~Chen and G.~K.~Leontaris,
``Perturbative moduli stabilisation in type IIB/F-theory framework'',
Eur. Phys. J. C \textbf{78} (2018) no.9, 766
[arXiv:1803.08941 [hep-th]].


\bibitem{Antoniadis:2019rkh}
I.~Antoniadis, Y.~Chen and G.~K.~Leontaris,
``Logarithmic loop corrections, moduli stabilisation and de Sitter vacua in string theory'',
JHEP \textbf{01} (2020), 149
[arXiv:1909.10525 [hep-th]].


\bibitem{Antoniadis:2020stf}
I.~Antoniadis, O.~Lacombe and G.~K.~Leontaris,
``Inflation near a metastable de Sitter vacuum from moduli stabilisation'',
Eur. Phys. J. C \textbf{80} (2020) no.11, 1014
[arXiv:2007.10362 [hep-th]].


\bibitem{Cribiori:2020use}
N.~Cribiori, G.~Dall'agata and F.~Farakos,
``Weak gravity versus de Sitter'',
JHEP \textbf{04} (2021), 046
[arXiv:2011.06597 [hep-th]].


\bibitem{Andriot:2021rdy}
D.~Andriot,
``Tachyonic de Sitter Solutions of 10d Type II Supergravities'',
Fortsch. Phys. \textbf{69} (2021) no.7, 2100063
[arXiv:2101.06251 [hep-th]].


\bibitem{Montero:2021otb}
M.~Montero, C.~Vafa, T.~Van Riet and G.~Venken,
``The FL bound and its phenomenological implications'',
JHEP \textbf{10} (2021), 009
[arXiv:2106.07650 [hep-th]].


\bibitem{DallAgata:2021nnr}
G.~Dall'Agata, M.~Emelin, F.~Farakos and M.~Morittu,
``The unbearable lightness of charged gravitini'',
JHEP \textbf{10} (2021), 076
[arXiv:2108.04254 [hep-th]].


\bibitem{Emelin:2022wft}
M.~Emelin,
``Obstacles for dS in Supersymmetric Theories'',
[arXiv:2206.01603 [hep-th]].


\bibitem{Jasinschi:1984cx}
R.~S.~Jasinschi and A.~W.~Smith,
``EFFECTIVE POTENTIAL IN N=1, d = 4 SUPERGRAVITY COUPLED TO THE VOLKOV-AKULOV FIELD'',
Phys. Lett. B \textbf{174} (1986), 183-185


\bibitem{Ellis:2013zsa}
J.~Ellis and N.~E.~Mavromatos,
``Inflation induced by gravitino condensation in supergravity'',
Phys. Rev. D \textbf{88} (2013) no.8, 085029
[arXiv:1308.1906 [hep-th]].


\bibitem{Alexandre:2013iva}
J.~Alexandre, N.~Houston and N.~E.~Mavromatos,
``Dynamical Supergravity Breaking via the Super-Higgs Effect Revisited'',
Phys. Rev. D \textbf{88} (2013), 125017
[arXiv:1310.4122 [hep-th]].


\bibitem{Ishikawa:2019pnb}
R.~Ishikawa and S.~V.~Ketov,
``Gravitino condensate in $N=1$ supergravity coupled to the $N=1$ supersymmetric Born-Infeld theory'',
PTEP \textbf{2020} (2020) no.1, 013B05
[arXiv:1904.08586 [hep-th]].


\bibitem{Alexandre:2014lla}
J.~Alexandre, N.~Houston and N.~E.~Mavromatos,
``Inflation via Gravitino Condensation in Dynamically Broken Supergravity'',
Int. J. Mod. Phys. D \textbf{24} (2015) no.04, 1541004
[arXiv:1409.3183 [gr-qc]]. 


\bibitem{Kehagias:2009zz}
A.~Kehagias,
``De Sitter vacua in simple extended supergravity'',
Fortsch. Phys. \textbf{57} (2009), 606-610


\bibitem{Polchinski:1983gv}
J.~Polchinski,
``Renormalization and Effective Lagrangians'',
Nucl. Phys. B \textbf{231} (1984), 269-295.


\bibitem{Kallosh:2022fsc}
R.~Kallosh, A.~Linde, T.~Wrase and Y.~Yamada,
``Goldstino condensation?,''
JHEP \textbf{08} (2022), 166
[arXiv:2206.04210 [hep-th]].


\bibitem{Volkov:1973ix}
D.~V.~Volkov and V.~P.~Akulov,
``Is the Neutrino a Goldstone Particle?'',
Phys. Lett. B \textbf{46} (1973), 109-110.


\bibitem{Clark:2000rv}
T.~E.~Clark and S.~T.~Love,
``The Akulov-Volkov Lagrangian, symmetry currents and spontaneously broken extended supersymmetry'',
Phys. Rev. D \textbf{63} (2000), 065012
[arXiv:hep-th/0007225 [hep-th]].


\bibitem{Cribiori:2016hdz}
N.~Cribiori, G.~Dall'Agata and F.~Farakos,
``Interactions of N Goldstini in Superspace'',
Phys. Rev. D \textbf{94} (2016) no.6, 065019
[arXiv:1607.01277 [hep-th]].


\bibitem{tHooft:1973alw}
G.~'t Hooft,
``A Planar Diagram Theory for Strong Interactions'',
Nucl. Phys. B \textbf{72} (1974), 461


\bibitem{Kraus:2002sa}
P.~Kraus and E.~T.~Tomboulis,
``Photons and gravitons as Goldstone bosons, and the cosmological constant'',
Phys. Rev. D \textbf{66} (2002), 045015
[arXiv:hep-th/0203221 [hep-th]].


\bibitem{deRham:2014zqa}
C.~de Rham,
``Massive Gravity'',
Living Rev. Rel. \textbf{17} (2014), 7
[arXiv:1401.4173 [hep-th]].


\bibitem{Folkerts:2013mra}
S.~Folkerts, C.~Germani and N.~Wintergerst,
``Massive spin-2 theories'',
[arXiv:1310.0453 [hep-th]].


\bibitem{Ferrara:2018wlb}
S.~Ferrara, A.~Kehagias and D.~L\"ust,
``Bimetric, Conformal Supergravity and its Superstring Embedding'',
JHEP \textbf{05} (2019), 100
[arXiv:1810.08147 [hep-th]].


\bibitem{Hawking:2001yt}
S.~W.~Hawking and T.~Hertog,
``Living with ghosts'',
Phys. Rev. D \textbf{65} (2002), 103515
[arXiv:hep-th/0107088 [hep-th]].


\bibitem{Farakos:2020wfc}
F.~Farakos, A.~Kehagias and N.~Liatsos,
``de Sitter decay through goldstino evaporation'',
JHEP \textbf{02} (2021), 186
[arXiv:2009.03335 [hep-th]].


\bibitem{Engelbrecht:2022aao}
L.~Engelbrecht, C.~R.~T.~Jones and S.~Paranjape,
``Supersymmetric Massive Gravity'',
JHEP \textbf{10} (2022), 130
[arXiv:2205.12982 [hep-th]].


\bibitem{Bandos:2016xyu}
I.~Bandos, M.~Heller, S.~M.~Kuzenko, L.~Martucci and D.~Sorokin,
``The Goldstino brane, the constrained superfields and matter in $ \mathcal{N}=1 $ supergravity'',
JHEP \textbf{11} (2016), 109
[arXiv:1608.05908 [hep-th]].






\bibitem{Lindstrom:1979kq}
U.~Lindstrom and M.~Rocek,
``CONSTRAINED LOCAL SUPERFIELDS,''
Phys. Rev. D \textbf{19} (1979), 2300-2303




\bibitem{Kapustnikov:1981de}
A.~A.~Kapustnikov,
``NONLINEAR REALIZATION OF EINSTEINIAN SUPERGRAVITY,''
Theor. Math. Phys. \textbf{47} (1981), 406-413


\bibitem{Farakos:2013ih}
F.~Farakos and A.~Kehagias,
``Decoupling Limits of sGoldstino Modes in Global and Local Supersymmetry,''
Phys. Lett. B \textbf{724} (2013), 322-327
[arXiv:1302.0866 [hep-th]]. 



\bibitem{Kallosh:2014wsa}
R.~Kallosh and T.~Wrase,
``Emergence of Spontaneously Broken Supersymmetry on an Anti-D3-Brane in KKLT dS Vacua'',
JHEP \textbf{12} (2014), 117
[arXiv:1411.1121 [hep-th]].


\bibitem{Bergshoeff:2015jxa}
E.~A.~Bergshoeff, K.~Dasgupta, R.~Kallosh, A.~Van Proeyen and T.~Wrase,
``$ \overline{\mathrm{D}3} $ and dS'',
JHEP \textbf{05} (2015), 058
[arXiv:1502.07627 [hep-th]].




\bibitem{Dudas:2015eha}
E.~Dudas, S.~Ferrara, A.~Kehagias and A.~Sagnotti,
``Properties of Nilpotent Supergravity,''
JHEP \textbf{09} (2015), 217
[arXiv:1507.07842 [hep-th]].




\bibitem{Bergshoeff:2015tra}
E.~A.~Bergshoeff, D.~Z.~Freedman, R.~Kallosh and A.~Van Proeyen,
``Pure de Sitter Supergravity,''
Phys. Rev. D \textbf{92} (2015) no.8, 085040
[erratum: Phys. Rev. D \textbf{93} (2016) no.6, 069901]
[arXiv:1507.08264 [hep-th]].







\bibitem{Dasgupta:2016prs}
K.~Dasgupta, M.~Emelin and E.~McDonough,
``Fermions on the antibrane: Higher order interactions and spontaneously broken supersymmetry'',
Phys. Rev. D \textbf{95} (2017) no.2, 026003
[arXiv:1601.03409 [hep-th]].






\bibitem{DallAgata:2016syy}
G.~Dall'Agata, E.~Dudas and F.~Farakos,
``On the origin of constrained superfields,''
JHEP \textbf{05} (2016), 041
[arXiv:1603.03416 [hep-th]].



\bibitem{Vercnocke:2016fbt}
B.~Vercnocke and T.~Wrase,
``Constrained superfields from an anti-D3-brane in KKLT'',
JHEP \textbf{08} (2016), 132
[arXiv:1605.03961 [hep-th]].






\bibitem{Kallosh:2016aep}
R.~Kallosh, B.~Vercnocke and T.~Wrase,
``String Theory Origin of Constrained Multiplets'',
JHEP \textbf{09} (2016), 063
[arXiv:1606.09245 [hep-th]].



\bibitem{Cribiori:2019hod}
N.~Cribiori, C.~Roupec, T.~Wrase and Y.~Yamada,
``Supersymmetric anti-D3-brane action in the Kachru-Kallosh-Linde-Trivedi setup,''
Phys. Rev. D \textbf{100} (2019) no.6, 066001
[arXiv:1906.07727 [hep-th]].


\bibitem{Bena:2020xrh}
I.~Bena, J.~Bl\r{a}b\"ack, M.~Gra\~na and S.~L\"ust,
``The tadpole problem'',
JHEP \textbf{11} (2021), 223
[arXiv:2010.10519 [hep-th]].


\bibitem{Plauschinn:2021hkp}
E.~Plauschinn,
``The tadpole conjecture at large complex-structure'',
JHEP \textbf{02} (2022), 206
[arXiv:2109.00029 [hep-th]].


\bibitem{Schnitzer:1974ji}
H.~J.~Schnitzer,
``Nonperturbative Effective Potential for Lambda phi**4 Theory in the Many Field Limit'',
Phys. Rev. D \textbf{10} (1974), 1800


\bibitem{Coleman:1974jh}
S.~R.~Coleman, R.~Jackiw and H.~D.~Politzer,
``Spontaneous Symmetry Breaking in the O(N) Model for Large N*'',
Phys. Rev. D \textbf{10} (1974), 2491


\bibitem{Kobayashi:1975ev}
M.~Kobayashi and T.~Kugo,
``On the Ground State of O(n)-Lambda phi**4 Model'',
Prog. Theor. Phys. \textbf{54} (1975), 1537


\bibitem{Abbott:1975bn}
L.~F.~Abbott, J.~S.~Kang and H.~J.~Schnitzer,
``Bound States, Tachyons, and Restoration of Symmetry in the 1/N Expansion'',
Phys. Rev. D \textbf{13} (1976), 2212





\bibitem{Kachru:2002gs}
S.~Kachru, J.~Pearson and H.~L.~Verlinde,
``Brane / flux annihilation and the string dual of a nonsupersymmetric field theory'',
JHEP \textbf{06} (2002), 021
[arXiv:hep-th/0112197 [hep-th]]. 


\bibitem{Gautason:2015tla}
F.~F.~Gautason, B.~Truijen and T.~Van Riet,
``The many faces of brane-flux annihilation'',
JHEP \textbf{10} (2015), 152
[arXiv:1505.00159 [hep-th]].


\bibitem{Mourad:2017rrl}
J.~Mourad and A.~Sagnotti,
``An Update on Brane Supersymmetry Breaking'',
[arXiv:1711.11494 [hep-th]].




\end{thebibliography}
\end{document}